\documentclass[epj]{svjour}

\usepackage{graphicx}
\usepackage{fancyhdr}

\def\mytitle{Discovery potential of the LHC for extended gauge symmetries} 
\def\myauthors{Gernot Krobath representing the ATLAS and CMS collaborations}  
\def\mytype{Parallel}
\def\mysession{Alternatives}

\def\mytitle{Discovery potential of the LHC for extended gauge symmetries} 
\def\myauthors{Gernot Krobath}    
\def\mytype{Contributed Talk}    
\def\mysession{Alternatives}

\begin{document}
\title{Discovery potential of the LHC for extended gauge symmetries}
\author{Gernot Krobath representing the ATLAS and CMS collaborations 
}                     

\institute{Fakult\"at f\"ur Physik, LMU M\"unchen, Am Coulombwall 1, 85748 Garching}
%
\date{}
\abstract{
Many models of physics beyond the Standard Model are based on extended gauge
symmetries and predict the existence of new heavy particles, often at the TeV scale.
Such particles include heavy W and Z bosons, doubly charged higgses, heavy majorana
neutrinos, leptoquarks and heavy fermions.  The discovery potential of the LHC, which will start in 2008, for various particles predicted in extended gauge theories will be described in this paper.
} 
\maketitle
\section{Introduction}
\label{intro}
The Standard Model is a very successful model, it describes the interactions of matter excluding gravity. Even though it has predicted many properties very well, the Standard Model has some limitations:
\begin{itemize}
\item 19 parameters have to be extracted from experiments
\item no unification of forces
\item fine tuning problems
\item no explanation for the existence of 3 generations of quarks and leptons
\end{itemize}
Extended gauge symmetries solve some of these problems. There are many models of physics beyond the Standard Model, which are based on extended gauge symmetries:

\subsection{Sequential Standard Model}
\label{ssm}
In the Sequential Standard Model (SSM) \cite{Dimo} it is assumed that there are additional heavy gauge bosons Z' and W', which have the same couplings as the Standard Model gauge bosons. The SSM is considered as a toy model by some people, but it is a very helpful benchmark for the experimentalists in the search for new heavy gauge bosons because the assumed interactions are well known.

\subsection{Left-Right Symmetric Models}
\label{LRSM}
In the Standard Model only left-handed leptons couple via the weak interaction to the W. This asymmetry is being restored in Left-Right Symmetric Models (LRSM) by introducing right-handed partners to all left-handed fermions \cite{Pati} \cite{Moha}, so there are also right-handed neutrinos; furthermore right-handed fermions couple weakly to the W$_R^{\pm}$ in LRSM. The smallest gauge group to implement a LRSM is  $SU(3)_C \otimes SU(2)_L  \otimes SU(2)_R \otimes U(1)_{B-L}$. The symmetry is intact above a mass mass scale $\Lambda_{LR}$=M$_{W_{R}}$ and is broken spontaneously at energies below this mass scale. LRSM introduce additional particles W$_{R}^{\pm}$, Z' and right-handed heavy majorana neutrinos N.

\subsection{Grand-Unified Theories}
\label{GUTs}

The superstring inspired E$_{6}$ Models \cite{Gur} \cite{Hew} are Grand-Unified Theories (GUTs). In GUTs the particles are put into larger symmetry groups and the electroweak and strong forces unify at energies a few orders of magnitude below the Planck scale. The exceptional Lie Group E$_{6}$ has long been considered as one of the favorite candidates for such a GUT gauge symmetry group. In the E$_{6}$ Models each generation of the Standard Model is placed in a \textbf{27} representation. The models predict many new particles, among them an additional heavy gauge boson Z' and leptoquarks.\\
There are many other GUTs besides the superstring inspired E$_{6}$ Models: SO(10) \cite{Fritzsch}, SU(6) \cite{Dashen} etc. Many GUTs predict the existence of Leptoquarks among many other new particles.

\section{Search For Particles Predicted In Extended Gauge Theories}
\label{PPIEGT}

\subsection{W'}
\label{W'}
The W' in the SSM is assumed to have the same couplings as the W of the Standard Model and the cross-section of the W' at the LHC is assumed to be the same as the W scaled by $(\frac{M_{W}}{M_{W'}})^{2}$ (SSM). The lower bound on the mass of the W' is 1 TeV \cite{CDFwprime} \cite{D0wprime}. 

The channel studied in ATLAS is W' $\rightarrow$ $\mu$ + $\nu_{\mu}$. The Standard Model backgrounds which have to be considered are: W $\rightarrow$ $\mu$ + $\nu_{\mu}$ + X, Z $\rightarrow$ $\mu$ + $\mu$ + X and jet production of QCD processes. The W' signal is a high energetic muon accompanied by missing energy. This allows an easy separation of W' and background reactions. 

\begin{figure}
\centering
\includegraphics[
height=0.4\textwidth,angle=0]{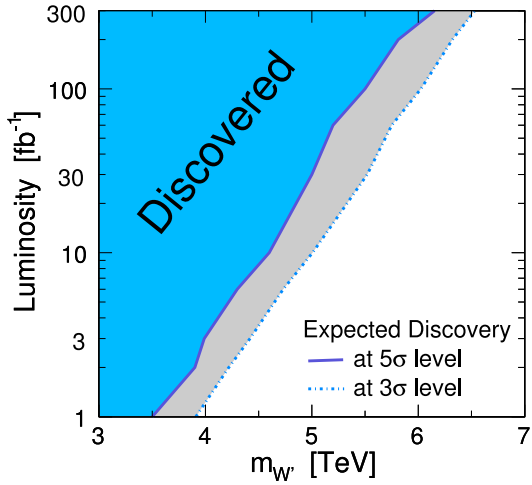}
\caption{Integrated luminosity needed to discover W' in CMS with a 3$\sigma$ and 5$\sigma$ significance respectively depending on the W' mass  \cite{CMSTDR}}
\label{Wprime}
\end{figure}

In CMS the same channel (W' $\rightarrow$ $\mu$ + $\nu_{\mu}$) is studied and the same background processes as in ATLAS were considered. In figure \ref{Wprime} the integrated luminosity needed for a 3$\sigma$ and 5$\sigma$ discovery respectively depending on the mass of the W' can be seen. For an integrated luminosity of 10 fb$^{-1}$, W' bosons of the SSM can be discovered or excluded up to a mass of 4.5-5 TeV, from an analysis of the muonic decay mode  \cite{CMSTDR}.

\subsection{Z'}
\label{Z}
\begin{figure}
\centering
\includegraphics[
height=0.5\textwidth,angle=0]{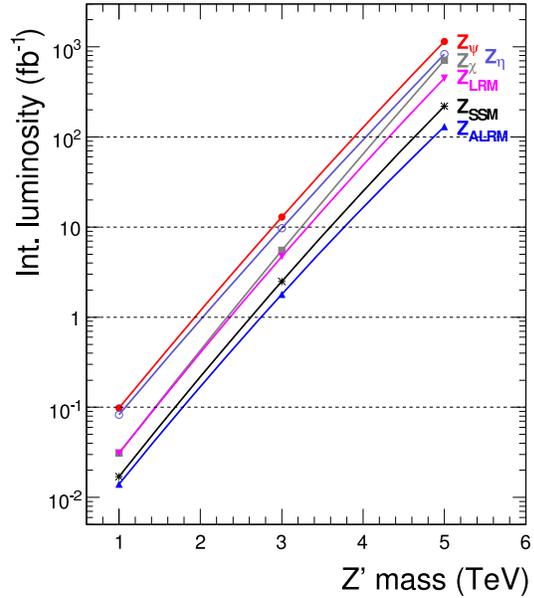}
\caption{Integrated luminosity needed to discover Z' in different models in CMS with a 5$\sigma$ significance depending on the Z' mass \cite{CMSTDR}}
\label{Zprime}
\end{figure}

The cross-section of the Z' at the LHC is assumed to be the same as the Z scaled by $(\frac{M_{Z}}{M_{Z'}})^{2}$ (SSM).
The mass region up to 850 GeV is excluded by Run II at the Tevatron \cite{ZTevatron}.
The studied channel of the Z' in CMS is Z' $\rightarrow$ $\mu^{+} + \mu^{-}$. The dominant and irreducible background is Z/$\gamma^{*}$ $\rightarrow$ $\mu^{+} + \mu^{-}$. The overall contribution from ZZ, ZW, WW, and $t\bar{t}$ is neglected because it was found to be at the level of only a few percent of the Drell-Yan background and can be further suppressed by signal-selection criteria with almost no reduction in signal efficiency. In figure \ref{Zprime} the expected luminosity needed to discover Z' in different models in this decay channel with 5$\sigma$ significance can be seen. Z$_{SSM}$ is the Z' within the sequential Standard Model; Z$_{\eta}$, Z$_{\psi}$ and Z$_{\chi}$ arise in E$_6$ (and SO(10)) GUT groups; Z$_{LRM}$ and Z$_{ALRM}$ are the Z' arising in the framework of the so-called "left-right" and "alternative left-right"models (g$_{R}$ = g$_{L}$ chosen). The discovery potential for the Z' in different models with an integrated luminosity of 1 fb$^{-1}$ and 5$\sigma$ significance can be seen in table \ref{Zdiscovery}.

\begin{table}
\begin{center}
\caption{Discovery potential for Z' of different models in CMS with an integrated luminosity of 1 fb$^{-1}$ and 5$\sigma$ significance \cite{CMSTDR}}
\label{Zdiscovery}       
\begin{tabular}{ll}
\hline\noalign{\smallskip}
Z' model & upper mass reach in TeV\\
\noalign{\smallskip}\hline\noalign{\smallskip}
 Z$_{SSM}$ & 2.6\\
Z$_{\eta}$ & 2\\
Z$_{\psi}$ & 1.95\\
Z$_{\chi}$ & 2.5\\
Z$_{LRM}$ & 2.5\\
Z$_{ALRM}$ & 2.7\\
\noalign{\smallskip}\hline
\end{tabular}
\end{center}
\end{table}

\subsection{Heavy Majorana Neutrinos, W$_{R}$}
\label{WR}
The LRSM incorporate three additional heavy gauge bosons W$_{R}^{\pm}$, Z' and the heavy right-handed Majorana neutrino states N. The Ns can be partner of light neutrino states and can provide the small but non-zero masses of the Standard Model neutrinos through the see-saw mechanism. It is assumed that g$_R$ = g$_L$.

\begin{figure}
\centering
\includegraphics[
height=0.45\textwidth,angle=0]{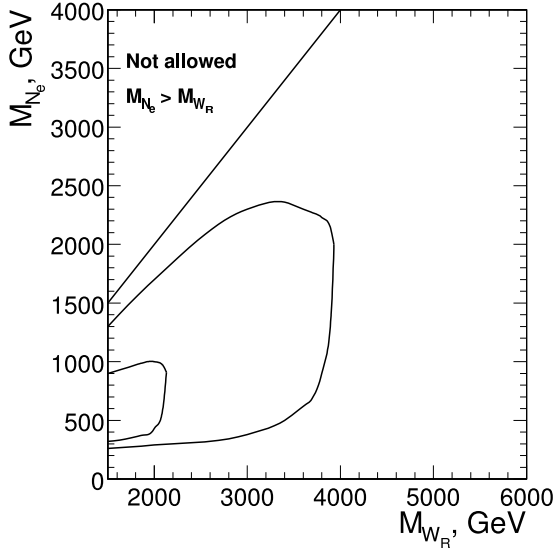}
\caption{Discovery potential for a 5$\sigma$ significance in CMS with an integrated luminosity of 1 fb$^{-1}$ (inner contour) and 30 fb$^{-1}$ (outer contour) depending on the M$_N$ and M$_{W_{R}}$. Everything inside the contour can be discovered \cite{CMSTDR}}
\label{MN2}
\end{figure}

The channel studied in CMS is pp $\rightarrow$ W$_R$ $\rightarrow$ e + N$_e$.
This channel has been chosen because the cross-section for this channel is 10 times higher than for pp $\rightarrow$ Z' $\rightarrow$ N$_e$ N$_e$. It is assumed that only the lightest N$_e$ is reachable at the LHC. The analysis is performed in the
M$_{W_{R}}$, M$_{N_{e}}$ parameter space. For the benchmark point considered the masses are: M$_{N_{e}}$ = 500 GeV and M$_{W_{R}}$ = 2000 GeV. The signal of W$_R$ decays is 2 leptons and 2 jets; the signal of N decays is one lepton and 2 jets. The main background processes are Z+jets and $t\bar{t}$ production. The 5$\sigma$ discovery contours can be seen in figure \ref{MN2} for an integrated luminosity of 1 fb$^{-1}$, 10 fb$^{-1}$ and 30 fb$^{-1}$. With 30 fb$^{-1}$ a 5$\sigma$ observation of W$_R$ and N$_e$ with masses up to 4 TeV and 2.4 TeV respectively can be achieved. The signal at the benchmark point (M$_{W_{R}}$ = 2 TeV and M$_{N_{e}}$ = 500 GeV) is observable already after one month of running at low luminosity \cite{CMSTDR}.

\subsection{Second Generation Scalar Leptoquarks}
\label{LQ}
\begin{figure}
\centering
\includegraphics[
height=0.25\textwidth,angle=0]{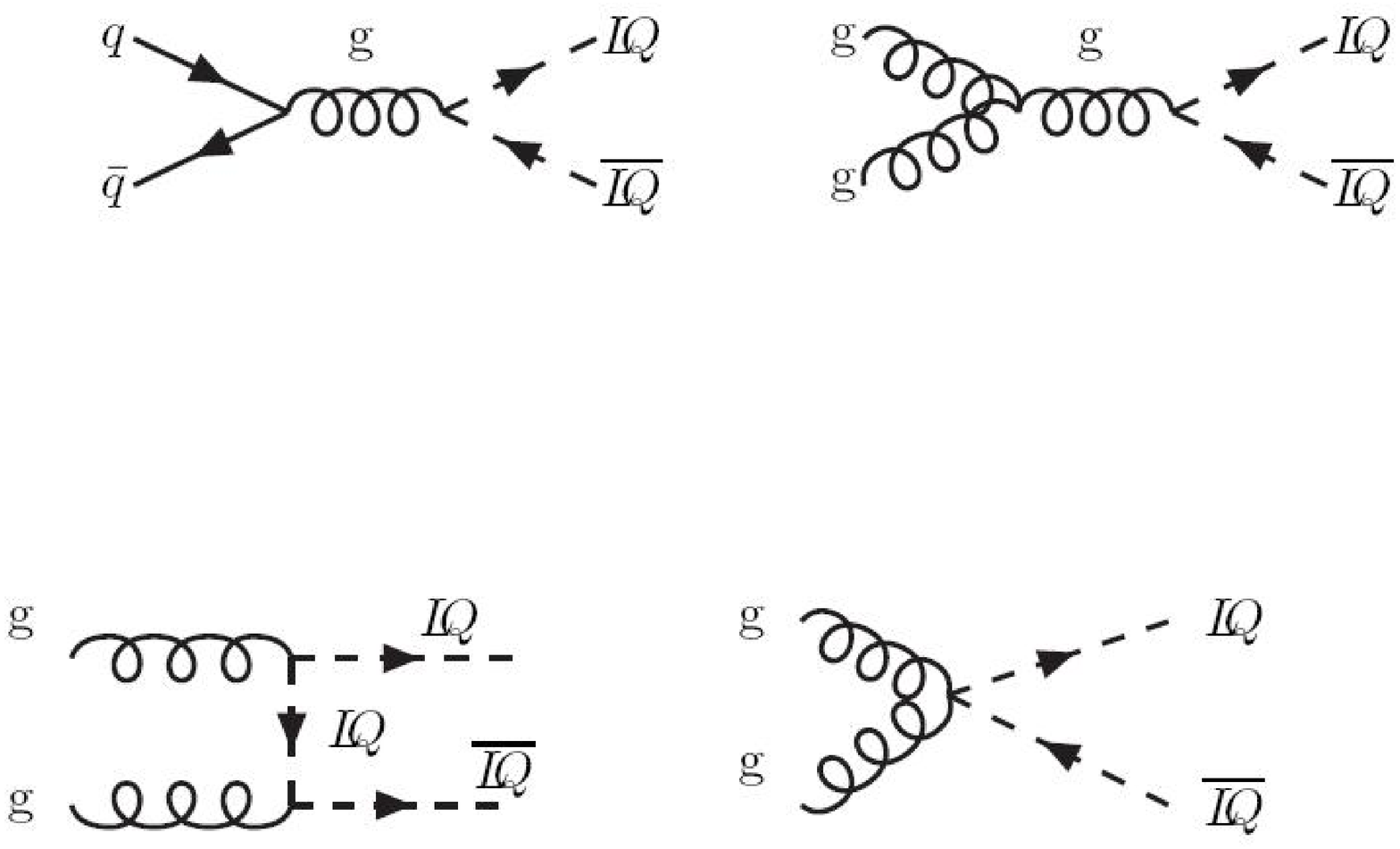}
\caption{Leptoquark pair production graphs}
\label{LQ-graph}
\end{figure}

Leptoquarks are particles which carry both lepton and baryon numbers. Leptoquark interactions conserve the lepton and baryon number separately. Since neither an excess of flavor changing neutral currents in the quark sector, nor in the charged lepton sector has been observed, it is assumed that Leptoquarks couple to only one generation of quarks and one generation of leptons of the Standard Model. Furthermore it is assumed that Leptoquark interactions are chiral otherwise rare decays, like the spin-suppressed decay $\pi^{-}$ $\rightarrow$ e$^{-}$ + $\bar\nu_{e}$, would be mediated by Leptoquarks. These assumptions lead to the minimum Buchm\"uller-R\"uckl-Wyler (mBRW) model  \cite{BRW}. There are 14 kinds of Leptoquarks in this model. Here only pair production of scalar Leptoquarks is considered (see figure \ref{LQ-graph}), since single production depends on the unknown Yukawa coupling (l-q-LQ coupling). The production of vector Leptoquarks depends on the spin of the incoming particles and the cross-section is significantly larger for vector Leptoquarks in general. Therefore scalar Leptoquark limits will be lower than vector Leptoquark limits. The pair production of scalar Leptoquarks depends only on the mass of the Leptoquarks. Here it is assumed that 100\% of the scalar Leptoquarks decay into LQ $\rightarrow$ $\l^{\pm}$ + q, i.e. branching fraction $\beta$ (LQ $\rightarrow$ $\l^{\pm}$ + j) = 1. Only second generation Leptoquarks are considered here, i.e. LQ $\rightarrow$ $\mu^{\pm}$ + q. There are also searches for first generation Leptoquarks and third generation Leptoquarks \cite{Mitsou}. The excluded mass limit for second generation scalar Leptoquarks with $\beta$ = 1 is $\sim$ 250 GeV \cite{Christiansen}. The cross-sections of scalar Leptoquarks at the LHC can be seen in table \ref{LQ-cross}. The main background processes are $t\bar{t}$ and Z/$\gamma^{*}$.

\begin{table}
\begin{center}
\caption{Scalar Leptoquark cross-sections at the LHC \cite{Kra2005}}
\label{LQ-cross}       
\begin{tabular}{ll}
\hline\noalign{\smallskip}
Leptoquark mass & cross-section in pb\\
\noalign{\smallskip}\hline\noalign{\smallskip}
300 GeV & 10.1\\
400 GeV & 2.24\\
600 GeV & 0.221\\
800 GeV & 0.0378\\
\noalign{\smallskip}\hline
\end{tabular}
\end{center}
\end{table}

The following requirements were imposed on each event in ATLAS:
2 muons with a reconstructed track in the inner detector and the muon detector with opposite charge and 2 jets were required in each event. Each jet must have a minimum transverse momentum of 25 GeV. Each muon has to have a transverse momentum exceeding 60 GeV. The mass of the dimuon system of the two selected muons has to be equal to or exceed 180 GeV. This cut is done to suppress background from the Z boson. These requirements and cuts are the same for all tested Leptoquark masses. Depending on the Leptoquark mass a cut on S$_{T}$ is made, where S$_{T}$ is the scalar sum of the transverse energies of the 2 selected muons and the 2 selected jets. The S$_{T}$-cut increases linearly with the tested Leptoquark mass: a minimum \mbox{S$_{T}$ of 350 GeV + m$_{LQ}^{tested}$} is required. These cuts were designed to achieve a maximum S/$\sqrt{B}$ for the $t\bar{t}$ and DY background. With 2 selected muons and 2 selected jets there are two possibilities to reconstruct the Leptoquark by adding the 4-vectors of 1 muon and 1 jet. Both combinations are calculated and the combination with the smallest difference between the two reconstructed Leptoquark masses is taken. The average of the masses of these two reconstructed Leptoquark masses is taken as reconstructed Leptoquark mass. Additionally a ``sliding mass window'' cut is applied on the events that are left after the cuts described so far. The reconstructed Leptoquark mass has to be in a mass range around the tested real Leptoquark mass where the range depends on the assumed real Leptoquark mass: the mass range is [0.75 $\times$ m$_{LQ}^{tested}$, 1.25 $\times$ m$_{LQ}^{tested}$]. Finally it is required that  $E\!\!\!\!/_{T}$/S$_{T}$ $<$ 0.1, where $E\!\!\!\!/_{T}$ is the missing transverse energy in an event. This cut is mainly made to suppress the (semi-)leptonic decays of the $t\bar{t}$ background.
In ATLAS an integrated luminosity of a few pb$^{-1}$ for the Leptoquark mass of 300 GeV and up to a few hundred pb$^{-1}$ for the Leptoquark mass of 800 GeV will be needed to exclude the tested Leptoquark mass with a 95\% confidence level.

\section{Conclusions}
\label{conclusions}
This paper presents a selection of analyses on particles predicted by extended gauge theories at \mbox{ATLAS} and CMS. The \mbox{ATLAS} and CMS experiments at the LHC provide a powerful tool to discover or exclude many particles predicted by extended gauge theories. Many of these particles can be discovered or excluded already in an early phase of the LHC.

\end{document}